\documentclass[12pt]{article}
\usepackage{amsfonts}
\usepackage{amsmath}
\usepackage{cite}

\csname @addtoreset\endcsname{equation}{section}
\newcommand{\tr}{{\rm Tr}\;}

\newcommand{\eq}{\begin{equation}}
\newcommand{\feq}{\end{equation}}
\newcommand{\eqn}{\begin{eqnarray}}
\newcommand{\feqn}{\end{eqnarray}}
\newcommand{\arr}{\begin{eqnarray*}}
\newcommand{\farr}{\end{eqnarray*}}

\def\tr{\mathop{\rm tr}\nolimits}
\def\det{\mathop{\rm det}\nolimits}    
\font\mybb=msbm10 at 12pt
\def\bb#1{\hbox{\mybb#1}}

\def\bR {\bb{R}}

\def\bC {\bb{C}}

\begin{document}

\begin{titlepage}
\begin{flushright}
UTF-447\\
IFUM-710-FT\\
hep-th/0203268
\end{flushright}
\vspace{.3cm}
\begin{center}
\renewcommand{\thefootnote}{\fnsymbol{footnote}}
{\Large \bf De~Sitter Gravity and Liouville
Theory}
\vskip 25mm
{\large \bf {Dietmar Klemm$^1$\footnote{dietmar.klemm@mi.infn.it}
and Luciano Vanzo$^2$\footnote{vanzo@science.unitn.it}}}\\
\renewcommand{\thefootnote}{\arabic{footnote}}
\setcounter{footnote}{0}
\vskip 10mm
{\small
$^1$ Dipartimento di Fisica dell'Universit\`a di Milano and\\
INFN, Sezione di Milano,
Via Celoria 16, I-20133 Milano.\\
\vspace*{0.5cm}

$^2$ Dipartimento di Fisica dell'Universit\`a di Trento and\\
INFN, Gruppo Collegato di Trento, Via Sommarive 14, I-38050 Povo (TN).
}
\end{center}
\vspace{2cm}
\begin{center}
{\bf Abstract}
\end{center}
{\small
We show that the spectrum of conical defects in three-dimensional
de~Sitter space is in one-to-one correspondence with the spectrum
of vertex operators in Liouville conformal field theory. The
classical conformal dimensions of vertex operators are equal
to the masses of the classical point particles in dS$_3$, that cause the
conical defect. The quantum dimensions instead are shown to coincide
with the mass of the Kerr-dS$_3$ solution computed with the
Brown-York stress tensor. Therefore classical de~Sitter gravity encodes the
quantum properties of Liouville theory. The equality of the
gravitational and the Liouville stress tensor provides a further
check of this correspondence. The Seiberg bound for vertex operators
translates on the bulk side into an upper mass bound for classical
point particles. Bulk solutions with cosmological event horizons
correspond to microscopic Liouville states, whereas those without horizons
correspond to macroscopic (normalizable) states.
We also comment on recent criticisms
by Dyson, Lindesay and Susskind, and point out that the contradictions
found by these authors may be resolved if the dual CFT is not able to
capture the thermal nature of de~Sitter space. Indeed we find that on the
CFT side, de~Sitter entropy is merely Liouville momentum, and thus has
no statistical interpretation in this approach.
}

\end{titlepage}

\section{Introduction}

Recently there has been an increasing interest in gravity on
de~Sitter (dS) spacetimes \cite{Balasubramanian:2001rb,Witten:2001kn}.
The motivation for this comes partially from recent astrophysical
data that point towards a positive cosmological constant,
but apart from phenomenological aspects,
it also remains an outstanding challenge to understand the role
of de~Sitter space in string theory, to clarify the microscopic
origin of de~Sitter entropy \cite{Gibbons:mu}\footnote{For microscopic
derivations of dS entropy based on the Chern-Simons formulation of
2+1 dimensional dS gravity, or on other approaches that are not directly
related to string theory, cf.~\cite{Maldacena:1998ih,Park:1998qk,
Banados:1998tb,Kim:1998zs,Lin:1999gf,Hawking:2000da,Govindarajan:2002ry}.},
and to study in which way the holographic
principle \cite{'tHooft:gx} is realized in the
case of de~Sitter gravity.
Whereas string theory on anti-de~Sitter spaces is known to have a
dual description in terms of certain superconformal field
theories \cite{Aharony:1999ti}, such an explicit duality emerging
from string theory is still missing for dS spacetimes.
Yet, based on \cite{Hull:1998vg}, where the first evidence for a dS/CFT
correspondence was given, and on related ideas that appeared
in \cite{Bousso:1999cb,Balasubramanian:2001rb},
Strominger proposed recently a holographic duality
relating quantum gravity on dS$_D$ to a Euclidean conformal
field theory residing
on the past (or alternatively future) boundary ${\cal I}^-$
(${\cal I}^+$) of dS$_D$ \cite{Strominger:2001pn}. Subsequently, this
correspondence was further explored in \cite{Klemm:2001ea,Cacciatori:2001un,
Balasubramanian:2001nb,Bousso:2001mw,Spradlin:2001nb,Leblond:2002ns,
Kabat:2002hj,Mazur:2001aa}.

In general, the conformal field theories in question seem to be
non-unitary. Indeed, if the bulk fields, to which CFT operators couple,
become sufficiently massive, the conformal weights of these operators
turn out to be complex \cite{Strominger:2001pn}. In a certain sense,
this is puzzling, because the bulk fields provide unitary representations
of the de~Sitter isometry group SO$(D,1)$ for arbitrarily large
masses \cite{Tagirov:1972vv}.

A full understanding of the proposed dS/CFT correspondence
seems to require an embedding of de~Sitter space
in string theory. De~Sitter solutions in ordinary supergravity
theories \cite{Hull:2001ii} break of course all supersymmetries,
and this makes it questionable how far one could trust a Maldacena-type
argument in this case. On the other hand, Hull's IIB$^{\ast}$
theory \cite{Hull:1998vg} admits a supersymmetric dS$_5$ $\times$ H$^5$
vacuum, but unfortunately the theory has ghosts.

While at present we do not yet have a satisfactory comprehension of
these issues, we can still try to get some insight by studying the simplest
explicit example of a dS/CFT correspondence, namely the one between
pure dS gravity in three dimensions, and Euclidean Liouville
theory \cite{Cacciatori:2001un}. This is the aim of the present paper.

We will see that Liouville field theory is able to capture many features
of 3d gravity with positive cosmological constant. First of all, it
reproduces correctly Strominger's central charge
$c = 3l/2G$ \cite{Strominger:2001pn}\footnote{Cf.~also \cite{Klemm:2001ea}
for alternative derivations of the central charge.} that appears in
the asymptotic symmetry algebra of dS$_3$ gravity. Furthermore, we show
that the spectrum of Liouville vertex operators is in one-to-one
correspondence with the spectrum of bulk (gravity) solutions.
While the classical conformal dimensions of these operators reproduce
exactly the mass of the point particle sources that are present on
the 3d gravity side, their quantum dimensions coincide with the mass
of the bulk solutions computed with the Brown-York stress tensor.
Classical de~Sitter gravity thus encodes the quantum properties of
Liouville theory. We shall see that features of Liouville theory,
like the Seiberg bound \cite{Seiberg:1990eb}, as well as the appearance
of macroscopic (normalizable) and microscopic (non-normalizable)
states \cite{Seiberg:1990eb} do all have an interpretation in 3d de~Sitter
gravity. In particular, bulk solutions with cosmological event horizons
correspond to microscopic Liouville states, whereas those without horizons
correspond to macroscopic states. Besides, complex conformal weights
occur naturally in Liouville theory, which is exactly what seems to be
required in a CFT dual to dS gravity.

The remainder of our paper is organized as follows:
In section \ref{KdSsol} we briefly review the Kerr-de~Sitter solution
in three dimensions, and compute the associated Brown-York boundary
stress tensor and the conserved charges like mass and angular momentum.
Furthermore, we discuss the thermodynamics of the solution, and argue
that a consistent thermodynamics can be formulated only in absence
of conical defects. We observe that the basic obstacle in
understanding thermodynamics is rooted in the absence of a zero
temperature state. In section \ref{OtherConf} we introduce other
configurations of de~Sitter gravity, just to remind ourselves and the
reader that Liouville field theory has to face many other problems
before one can say it is a successful description of quantum
three-dimensional gravity. 
In section \ref{dSLiou}, starting from the Liouville
action obtained in \cite{Cacciatori:2001un}, we compute the central charge
and show that it coincides with the one obtained in \cite{Strominger:2001pn}.
The Liouville field $\Phi$ corresponding to the Kerr-dS$_3$ solution is
then determined. This leads to a relation between Liouville vertex operators
$e^{\alpha\Phi}$ and bulk solutions. We show that the Liouville stress
tensor equals the Brown-York stress tensor of dS$_3$ gravity, and that
the quantum dimensions of vertex operators reproduce exactly the masses
of the bulk solutions. After that, it is shown what normalizable states
and the Seiberg bound in Liouville theory correspond to on the bulk side.
In section \ref{thermal}, we comment on recent criticisms
by Dyson, Lindesay and Susskind \cite{Dyson:2002nt}, and point out that the
contradictions found by these authors may be resolved if the dual CFT
is not able to capture the thermal nature of de~Sitter space.
Indeed we find that on the CFT side, de~Sitter entropy is merely Liouville
momentum, and thus has no statistical interpretation in this approach.
We conclude with some final remarks.

\section{The Kerr-de~Sitter solution}

\label{KdSsol}

Consider the Einstein-Hilbert action with positive cosmological
constant $\Lambda = l^{-2}$,

\begin{equation}
I = \frac{1}{16\pi G}\int_{{\cal M}}d^3x \sqrt{-g}\left(R - \frac{2}{l^2}
    \right) + \frac{1}{8\pi G}\int_{\partial {\cal M}}d^2x \sqrt{\gamma} K\,,
    \label{EHaction}
\end{equation}

where we included the Gibbons-Hawking
boundary term necessary to have a well-defined variational principle.
$K$ is the trace of the extrinsic curvature $K_{\mu\nu} = -\nabla_{\left(
\mu\right.}n_{\left.\nu\right)}$ of the spacetime boundary $\partial {\cal M}$,
with $n^{\mu}$ denoting the outward pointing unit normal.
$\gamma$ is the induced metric on the boundary.

The equations of motion following from (\ref{EHaction}) admit
the Kerr-de~Sitter solution given by

\begin{equation}
ds^2 = -N^2 dt^2 + N^{-2} dr^2 + r^2(d\phi + N^{\phi}dt)^2\,,
\label{KdS}
\end{equation}

with the lapse and shift functions

\begin{equation}
N^2 = \mu - \frac{r^2}{l^2} + \frac{16G^2J^2}{r^2}\,, \qquad
N^{\phi} = \frac{4GJ}{r^2}\,.
\end{equation}

(\ref{KdS}) describes the gravitational field of a point particle
with spin $J$, and mass related to the parameter $\mu$ (cf.~below).
The Kerr-dS spacetime has a cosmological event horizon located
at\footnote{The horizon exists for every $\mu \in \bR$ if $J \neq0$, and
for positive $\mu$ if $J = 0$.}

\begin{equation}
r = r_+ = \frac{l}{2}(\sqrt{\tau} + \sqrt{\bar{\tau}})\,,
\end{equation}

where we defined

\begin{equation}
\tau = \mu - \frac{8GJi}{l}\,. \label{tau}
\end{equation}

The Bekenstein-Hawking entropy, temperature, and angular velocity of
the horizon are given respectively by

\begin{eqnarray}
S &=& \frac{2\pi r_+}{4G} = \frac{\pi l}{4G}(\sqrt{\tau} +
      \sqrt{\bar{\tau}})\,, \nonumber \\
2\pi lT &=& \frac{r_+}{l} + \frac{16G^2J^2l}{r_+^3}\,, \label{STOmega} \\
\Omega &=& -\frac{4GJ}{r_+^2}\,. \nonumber
\end{eqnarray}

In order to compute the mass and angular momentum of
the spacetime (\ref{KdS}) one can proceed as in
\cite{Gibbons:mu}, and integrate the Killing identity in the static patch
from $r=0$ to $r=r_+$. In doing this, one has to take into account
the delta function sources at $r=0$. An alternative approach,
which we will pursue here,
makes use of the conserved stress tensor \cite{Brown:1992br}
associated to the boundary $\partial{\cal M}$. In our setting,
$\partial {\cal M}$ is the past boundary ${\cal I}^-$ ($r \to \infty$)
of the conformal
compactification of the Kerr-dS$_3$ spacetime, and the stress tensor
derived from the action (\ref{EHaction}) reads \cite{Strominger:2001pn}

\begin{equation}
T^{\mu\nu} = \frac{1}{8\pi G}\left[K^{\mu\nu} - K \gamma^{\mu\nu}
             - \frac 1l \gamma^{\mu\nu}\right]\,,
             \label{stresstensor}
\end{equation}

where the last term comes from a surface counterterm added to the action
(\ref{EHaction}) in order to cancel divergences \cite{Strominger:2001pn}.

For the metric (\ref{KdS}), a straightforward calculation yields for
$r \to \infty$

\begin{equation}
T_{tt} = -\frac{\mu}{16\pi Gl}\,, \qquad T_{t\phi} = \frac{J}{2\pi l}\,,
\qquad T_{\phi\phi} = \frac{\mu l}{16\pi G}\,. \label{stresstensKdS}
\end{equation}

Note that (\ref{stresstensKdS}) is traceless, $h^{\mu\nu}T_{\mu\nu} = 0$,
where $h_{\mu\nu}dx^{\mu}dx^{\nu} = dt^2 + l^2 d\phi^2$ denotes
the CFT metric. In complex coordinates $z = \phi + it/l$, the
energy-momentum tensor reads

\begin{equation}
T_{zz} \equiv T(z) = \frac{l\tau}{32\pi G}\,, \qquad
T_{\bar z\bar z} \equiv \overline{T}(\bar z) = \frac{l\bar \tau}{32\pi G}\,,
\qquad T_{z\bar z} = 0\,.
\end{equation}

For later use, we also need the stress tensor on the plane. One can
pass from the cylinder to the plane (with coordinates $w, \bar w$)
by setting

\begin{equation}
w = e^{iz}\,. \label{cylplane}
\end{equation}

This leads to

\begin{equation}
T(w) = \frac{1 - \tau}{4\pi\gamma^2 w^2}\,, \qquad
\overline{T}(\bar w) = \frac{1 - \bar{\tau}}{4\pi\gamma^2 \bar w^2}\,, \qquad
T_{w\bar w} = 0\,, \label{stresstensw}
\end{equation}

where the $\tau$- and $\bar{\tau}$-independent parts come from the
Schwartzian derivatives.

To each bulk Killing vector field $\xi$ one can associate a
conserved charge $Q_{\xi}$ \cite{Brown:1992br},

\begin{equation}
Q_{\xi} = \int_0^{2\pi} T_{\mu\nu} u^{\mu}\xi^{\nu} \sqrt{\sigma}d\phi\,,
\end{equation}

where $u = (-N^2)^{-1/2}(\partial_t - N^{\phi}\partial_{\phi})$ is the
unit normal to the surface $\Sigma_t$ of constant
$t$ in $\partial{\cal M}$, and $\sigma$ denotes the induced metric
on $\Sigma_t$. One then gets for the mass $Q_{\partial_t}$ and angular
momentum $Q_{\partial_{\phi}}$

\begin{equation}
Q_{\partial_t} = -\frac{\mu}{8G} =: M\,, \qquad Q_{\partial_{\phi}} = J\,,
\label{MJ}
\end{equation}

confirming that the parameter $J$ represents the angular momentum
of the system.

Consider now the case of vanishing angular momentum, $J=0$.
By the coordinate transformation

\begin{eqnarray}
r &=& \sqrt{\mu}l\cosh\tau\sin\theta\,, \nonumber \\
t &=& \frac{l}{\sqrt{\mu}}{\mathrm{artanh}}\left(\frac{\tanh\tau}
      {\cos\theta}\right)\,, \label{transfglobal} \\
\phi &=& \frac{\varphi}{\sqrt{\mu}}\,, \nonumber
\end{eqnarray}

the metric can be cast into the form

\begin{equation}
ds^2 = -l^2 d\tau^2 + l^2\cosh^2\tau (d\theta^2 + \sin^2\theta d\varphi^2)\,.
\label{metrglobal}
\end{equation}

(\ref{transfglobal}) is valid inside the static patch $0 \le r < r_+$.
For $r > r_+$ one has to use instead

\begin{equation}
t = \frac{l}{\sqrt{\mu}}{\mathrm{artanh}}(\coth\tau\cos\theta)\,,
\end{equation}

and the transformation formulas for $r$ and $\phi$ remain unchanged.

(\ref{metrglobal}) describes dS$_3$ as a contracting and expanding
two-sphere, with global coordinates $\tau, \theta, \varphi$.
We have to take into account however that $\varphi$ is identified
modulo $2\pi\sqrt{\mu}$, so one has a deficit angle $2\pi(1-\sqrt{\mu})$
for $\mu < 1$, and an excess angle $2\pi(\sqrt{\mu}-1)$ for $\mu > 1$.
One might assume that excess angles lead to inconsistencies. Yet, as
we shall see in section \ref{dSLiou}, excess angles imply negative
conformal weights of the corresponding vertex operators in the dual
Liouville field theory. As is well-known, this is
not a problem in Liouville theory, where the spectrum of conformal
dimensions is not bounded from below \cite{Seiberg:1990eb}.

Now we would like to discuss an important feature of the solutions
so far discussed. The Euclidean section of the spinless defects
suggests that Euclidean time is a periodic variable with
period $\beta = 2\pi l/\sqrt{\mu}$. The Wightman functions will share
this periodicity, so an Unruh detector will register a
local temperature $T_{\mathrm{loc}}=(-g_{tt})^{-1/2}\beta^{-1}$. However, this
does not necessarily mean that there exists a thermal equilibrium state
regular everywhere, because the metric is still singular at the
conical defect located in $r=0$ (this will produce divergences in the
stress tensor for matter at any temperature, near $r=0$) and there is
no possibility to obtain a Euclidean metric that is everywhere
nonsingular\footnote{This is somehow reminiscent of the higher-dimensional
case, where one has two horizons, a cosmological one and a black hole
horizon.}.\\ 
Anyway, thinking for a while in terms of temperature, one may
note that there is no zero temperature state in this case, since the
metric (\ref{KdS}) (with $J=0$) describes actually a cylindrical universe
expanding (or contracting) from (to) a wirelike singularity. The
solution is either timelike and null geodesically incomplete in the
past or in the future. The future incomplete solution can be written
in the form

\[
ds^2 = -dT^2+e^{-2T/l}(dR^2+l^2d\phi^2)\,,
\]

and the other has the scale factor with positive exponent. 
The non-existence of a zero temperature state
makes it hard to make sense of a statistical partition function, like
the familiar $\tr\exp(-\beta H)$.

We will now gain some evidence in favour of a non-thermal
interpretation of the partition function for de~Sitter gravity, by
computing the Euclidean action

\[
I = -\frac{1}{16\pi G}\int_{{\cal M}}d^3x \sqrt{-g}(R - 2l^{-2})
\]

for the solution (\ref{KdS}) with $J=0$. The action is not really
well-defined, so we cut a small disk around $r=0$ with radius $\epsilon$
and compute $I$ as a sum 
of two contributions, one from the disk and the other from the
complementary region. At the end we let $\epsilon\to 0$. The bulk
contribution trivially has a limit as $\epsilon\to 0$, which is

\begin{equation}
I_0=-\frac{\pi l\sqrt{\mu}}{2G}\,.
\label{microcan}
\end{equation}

The disk contribution can be calculated using the Gauss-Bonnet
theorem: the lapse function near $r=0$ is $\sqrt{\mu}$, the
three-dimensional scalar curvature reduces to the two-dimensional
curvature on the disk, so Gauss-Bonnet gives

\[
I_{\mathrm{disk}}=-\frac{\pi l}{2G}(1-\sqrt{\mu})\,.
\]

Since $I = I_0 + I_{\mathrm{disk}}$, the $\mu$-dependent terms cancel and
we are left with the Euclidean action of pure de Sitter space! This
indicates that the static, structureless pointlike defect carries no
entropy, which entirely comes from the cosmological horizon. Indeed,
the disk contribution to the action can be written as
$-\beta_{\mathrm{dS}}m_{\mathrm{class}}$, where
$\beta_{\mathrm{dS}} = 2\pi l$ is the
inverse temperature of de~Sitter space
and $m_{\mathrm{class}} = (1-\sqrt{\mu})/4G$ is the classical mass
of the point defect (see below). This also suggests to take $I=I_0$,
i.e. one deletes the singular point $r=0$ from the manifold to obtain a
regular one. Interpreted as a thermal partition function, this gives an
entropy which is twice as large as the Bekenstein-Hawking
entropy. Interpreted as a microcanonical partition function instead,
it gives an entropy

\[
S=\frac{\pi l\sqrt{\mu}}{2G}=\frac{A}{4G}\,.
\]

To check the internal consistency of the microcanonical
interpretation, we note that if we knew the mass the temperature could
be derived from the Gibbons-Hawking differential mass
formula \cite{Gibbons:mu} in de~Sitter space, which states that

\[
\frac{dS}{d(-M)}=\frac{2\pi}{\kappa} = \beta\,,
\]

where $\kappa$ is the surface gravity of the cosmological
horizon and $M$ is the mass within the horizon (so $-M$ is the mass
outside the horizon). Considering the mass (\ref{MJ}) we find indeed
the correct relation

\[
\frac{dS}{d(-M)} = \frac{2\pi l}{\sqrt{\mu}}\,.
\]

The canonical partition function $I_c$ is a different object, and
is defined as the Legendre transform of $I_0$,

\[
I_c = I_0 - M\frac{dI_0}{dM} = I_0 - \beta M = -\frac{\pi^2l^2}{2G}\,
      \beta^{-1}\,.
\]

The internal energy in the canonical ensemble is then
$U = \partial_{\beta}I_c = -M$ and the entropy is $A/4G$
again\footnote{For similar conclusions in the case of the
Schwarz\-schild-de~Sitter solution in four dimensions
cf.~\cite{Teitelboim:2002cv}.}. It is worth noting that by adding to $I_c$ the
term $\beta E_c$, where $lE_c=-(c+\tilde{c})/24$ is the Casimir
energy, one obtains a modular invariant partition function, invariant
under $\beta\to 4\pi^2l^2/\beta$.

Comparing with AdS$_3$, the situation is clear: there the action
is infinite, but we have a zero temperature state to compare with, 
the difference between the two actions is finite and describes
correctly the semi-classical thermodynamics of AdS$_3$ black
holes. In the de~Sitter case, the action is already finite but there is no
zero temperature state to compare with, so the action has no simple
thermodynamical interpretation. Instead, it can be interpreted
consistently as a microcanonical partition function. In this case,
$\exp(-I)$ is directly related to the probability of a de Sitter
configuration with given energy (see, e.g. \cite{Weinberg:1988cp} and
references therein).

Below we will propose an alternative description of de~Sitter
entropy in Liouville field theory. We shall see that $S$ is essentially
the Liouville momentum, and has thus no statistical interpretation in
this approach.

\section{Other configurations}

\label{OtherConf}

The Kerr-de Sitter solution is of course not unique,
three-dimensional gravity with positive cosmological constant
$\Lambda = l^{-2}$ having a lot of
other solutions. The only degree of freedom of the spherical sector
of 3d gravity is the total volume $\int\sqrt{g^{(2)}}$ of the
space. For different topological sectors there is more than this, the
moduli of the spatial metric being a complete set of configuration
variables \cite{Martinec:1984fs,Witten:1988hc,Moncrief:1989dx,Ezawa:1993ti}.
We present here the general vacuum
solution with toroidal topology, and confirm that this also leads to a
central charge $3l/2G$. This will amount to show that the asymptotic
behaviour at past or future infinity is locally the same as for de
Sitter gravity.

In a gauge where the spatial metric depends only on time,

\[
ds^2=-dt^2+g_{ij}(t)dx^idx^j\,, \qquad x^i = (x,y)\,,
\]

the field equations (in a self-explanatory matrix notation) are
  
\begin{equation}
\partial_t\tr(g^{-1}\dot{g})+\frac{1}{2}\tr(g^{-1}\dot{g}g^{-1}\dot{g})=
4\Lambda\,,
\label{rtt1}
\end{equation}

\begin{equation}
\partial_t(g^{-1}\dot{g})+\frac{1}{2}g^{-1}\dot{g}\tr(g^{-1}\dot{g})=
4\Lambda\,.
\label{rij1}
\end{equation}

One consequence is the identity expressing the Hamiltonian constraint of
general relativity,

\begin{equation}
\tr(g^{-1}\dot{g})\tr(g^{-1}\dot{g})-\tr(g^{-1}\dot{g}g^{-1}\dot{g})=
8\Lambda\,.
\label{iden}
\end{equation}

It is readily verified that (\ref{iden}) is equivalent to the statement
that $\det(g^{-1}\dot{g})=4\Lambda$. In this case

\[
S = \frac l2 g^{-1}\,\dot{g}\in {\mathrm{SL}}(2,\bR)\,.
\]

The Einstein equations admit then an SL$(2,\bR)$ $\times $
SL$(2,\bR)$ symmetry, $g\to UgV^{-1}$, under which the matrix $S$
transform as $S\to VSV^{-1}$. Since we are assuming that space is a
torus, the general solution must depend on four integration constants,
corresponding to the two moduli of a torus and their conjugate
momenta. We can encode these constants into the periods of the spatial
variables, which we assume to be identified according to

\[
(x,y)\simeq (x+q_1,y+q_2)\simeq (x+q_3,y+q_4)\,.
\]

It is not hard to verify that the general solution of the field
equations is (almost, but not quite)

\begin{equation}
ds^2 = -dt^2 + \cosh^2(t/l)dx^2+\sinh^2(t/l)dy^2\,,
\label{soluzione}
\end{equation}

with the given identifications for the pair $(x,y)$. There is also a
special solution

\begin{equation}
ds^2 = -dt^2+e^{\pm2t/l}|dx+\tau dy|^2\,, \qquad \tau = \tau_1 + i\tau_2\,,
\label{special}
\end{equation}

this time with $(x,y)$ ranging over the unit square. $\tau$ with
$\tau_2>0$ is the modulus of the torus. Passing to the new variables
$(\theta_1,\theta_2)$ given by $x=q_1\theta_1+q_2\theta_2$,
$y=q_3\theta_1+q_4\theta_2$, the metric (\ref{soluzione}) can also be
cast in the form

\[
ds^2 = -dt^2 + (q_1^2\cosh^2(t/l)+q_3^2\sinh^2(t/l))
|d\theta_1 + \tau(t)d\theta_2|^2\,,
\]

where $\tau(t)$ is a time dependent modulus which goes over to a
constant $\tau$ in the limits $t\to\pm\infty$. The special
solution (\ref{special}) is thus the asymptotic limit of the general
solution. In both cases the metric approaches

\[
ds^2 = -dt^2+\frac{1}{4}e^{-2t/l}dzd\bar{z}
\]

asymptotically at $t \to -\infty$.
Note also that the solutions either expand from a wirelike singularity
to an infinite volume torus, or either collapse from an infinite torus
to a wirelike singularity. So there is only one conformal boundary in
this case. This observation is pertinent to Strominger's proposal that
there is actually only one CFT for three-dimensional
de~Sitter gravity \cite{Strominger:2001pn}.

The given asymptotic form is locally the same as the de~Sitter case
(globally not, of course, the conformal Killing group of the torus is
two-dimensional), so Strominger's analysis can be applied: the
central charge must be given by $3l/2G$. The boundary stress tensor
(\ref{stresstensor}) has vanishing trace, which is consistent with the
fact that the scalar curvature of the torus vanishes.

Now we observe that there cannot be globally static solutions, because
the Hamiltonian constraint would give the curvature of the torus 
as twice the cosmological constant. This contradicts the fact that the
Euler characteristic vanishes. There should also be no locally static
solutions (we have not proved this), as the space volume always end or
begins in a wirelike singularity\footnote{There is, however, a
cosmological horizon, so the possibility exists for a coordinate
transformation to a locally static region.}. The torus Liouville field
theory must then be compared with the dynamics, a challenge for the 
dS$_3$/CFT$_2$ correspondence. The difficulty is that the Liouville
equation appears as the Hamiltonian constraint for Einstein
gravity for a spatial metric which is conformal to a constant curvature
metric. Thus, classically, the correspondence would mean that the
dynamics reduces to the solution of the constraint equations. It was
pointed out by Witten \cite{Witten:1988hc} that there is a gauge, in the
Chern-Simon formulation of 3d gravity, in which the field equations
reduce indeed to the constraints, but such a gauge cannot be applied 
to Einstein equations.

\section{dS$_3$ gravity and Liouville theory}

\label{dSLiou}

In \cite{Cacciatori:2001un} it was shown that the asymptotic dynamics
of pure de~Sitter gravity in three dimensions is described by
Euclidean Liouville field theory. This correspondence was established
by writing dS$_3$ gravity as an SL$(2,\bC)$ Chern-Simons theory,
with action

\begin{equation}
I = \frac{is}{4\pi}\int{\mathrm{Tr}}(A\wedge dA + \frac 23
    A\wedge A\wedge A) -
    \frac{is}{4\pi}\int{\mathrm{Tr}}(\bar{A}\wedge d\bar{A} + \frac 23
    \bar{A}\wedge \bar{A}\wedge \bar{A})\,, \label{CSaction}
\end{equation}

where, in the conventions of \cite{Cacciatori:2001un},

\begin{equation}
s = -\frac{l}{4G}\,,
\end{equation}

and

\begin{equation}
A = A^a\tau_a = \left(\omega^a + \frac il e^a\right)\tau_a\,, \qquad
\bar A = {\bar A}^a\tau_a = \left(\omega^a - \frac il e^a\right)\tau_a\,,
\quad a = 0,1,2\,, \label{CSconn}
\end{equation}

where $e^a$ is the dreibein,
$\omega^a = \frac 12 \epsilon^{abc}\omega_{bc}$
the spin connection, and the $\tau_a$ are the SL$(2,\bC)$
generators given in \cite{Cacciatori:2001un}.

Chern-Simons theory is known to reduce to a WZNW model in presence of
a boundary \cite{Elitzur:1989nr}. The boundary conditions for
asymptotically past de~Sitter spaces \cite{Strominger:2001pn} provide then
the constraints for a Hamiltonian reduction from the WZNW model
to Liouville field theory.

The Liouville action obtained in \cite{Cacciatori:2001un} reads

\begin{equation}
I = -\frac{1}{8\pi G}\int d\phi dt \left[\frac 12 \partial_z \Phi
    \partial_{\bar z} \Phi + 2\exp\Phi\right]\,, 
\end{equation}

where we used the complex coordinate $z = \phi + it/l$.
In order to conform with the conventions of \cite{Seiberg:1990eb},
we rescale $\Phi \to \gamma\Phi$, with $\gamma = \sqrt{8G/l}$.
This leads to the action

\begin{equation}
I = -\frac{1}{4\pi}\int \sqrt h d^2x \left[\frac 12 h^{ij}\partial_i
    \Phi\partial_j\Phi + \frac{\lambda}{2\gamma^2}\exp (\gamma\Phi)\right]\,,
\label{Liouvilleact}
\end{equation}

where the "cosmological constant" $\lambda$ is given by
$\lambda = 16l^{-2}$, and the CFT metric on the cylinder is
$h_{ij}dx^i dx^j = dt^2 + l^2 d\phi^2$. (\ref{Liouvilleact}) coincides,
up to the overall sign, with the expression given in
\cite{Seiberg:1990eb}. There are some subtleties concerning
this sign difference. First of all, when deriving (\ref{Liouvilleact}),
one starts from a Lorentzian gravitational (or Chern-Simons) action,
end ends up with a Euclidean action. This usually involves an additional
minus sign, which would eliminate the minus in front of (\ref{Liouvilleact}).
Presumably these sign ambiguities are also related to the different
definitions of the stress tensor used in \cite{Strominger:2001pn,Klemm:2001ea}
and \cite{Balasubramanian:2001nb}, that differ by an overall sign.
The former implies a positive central charge, whereas
the latter leads to $c=-3l/2G$ \cite{Balasubramanian:2001nb}.
Some discussion of these ambiguities can be found
in \cite{Balasubramanian:2001nb}.
It would certainly be desirable to understand these better. For the time being,
we use the plus sign in front of (\ref{Liouvilleact}). It will turn
out that this leads in fact to results that make sense
physically. First of all, we can check that the value of the central
charge in Liouville theory matches perfectly with that computed from
the gravitational side. We have only to remember that
the action (\ref{Liouvilleact}) was obtained in a gauge with flat
fiducial metric $h_{ij}$: changing to a
conformal gauge, $h_{ij} \to g_{ij} = e^{2u}h_{ij}$, the
Liouville field
changes from $\gamma\Phi$ to $\gamma\Phi - 2u$. Substituting into
Eq.~(\ref{Liouvilleact}) (with a plus sign in front) and observing
that $R = -e^{-2u}\nabla^2u$, one
obtains the action (modulo a field independent term)

\begin{equation}
I = \frac{1}{4\pi}\int \sqrt g d^2x \left[\frac 12 g^{ij}\partial_i
    \Phi\partial_j\Phi + \frac{\lambda}{2\gamma^2}\exp (\gamma\Phi)
+\frac{Q}{2}\,R\Phi\right]\,,
\label{Liouvilleact1}
\end{equation}

where

\[
Q = \sqrt{\frac{l}{2G}}
\]

is the background charge. As is well-known, this theory has a
classical central charge $c=3Q^2$, which is equal to the de Sitter
charge $3l/2G$. This is positive, as it should be, when one uses the
conventions of \cite{Strominger:2001pn,Klemm:2001ea}.

As a further consistency check, we will show below
that the Liouville stress tensor reproduces
the gravitational energy-momentum tensor (\ref{stresstensor}), which
was defined with the conventions of \cite{Strominger:2001pn,Klemm:2001ea}.
Note also that in this way we obtain for the classical Liouville central
charge

\begin{equation}
c = \frac{12}{\gamma^2}\,,
\end{equation}

confirming the classical relation of Liouville theory $Q=2/\gamma$. 
For the Kerr-dS$_3$ solution (\ref{KdS}), the Chern-Simons
connection (\ref{CSconn}) has the asymptotic behaviour for $r \to \infty$

\begin{eqnarray}
A_r &=& \left(\begin{array}{cc} \frac{1}{2r} & 0 \\ 0 & -\frac{1}{2r}
      \end{array}\right)\,, \qquad
A_z = \left(\begin{array}{cc} 0 & \frac{il\tau}{4r} \\ \frac{ir}{l} & 0
      \end{array}\right)\,, \qquad
A_{\bar z} = 0\,, \nonumber \\
\bar{A}_r &=& \left(\begin{array}{cc} -\frac{1}{2r} & 0 \\ 0 & \frac{1}{2r}
      \end{array}\right)\,, \qquad
\bar{A}_{\bar z} = \left(\begin{array}{cc} 0 & -\frac{ir}{l} \\
                   -\frac{il\bar{\tau}}{4r} & 0 \end{array}\right)\,, \qquad
\bar{A}_z = 0\,.
\end{eqnarray}

Following \cite{Cacciatori:2001un}, it is straightforward to compute
the corresponding Liouville solution

\begin{equation}
e^{\gamma\Phi} =  \frac{\tau\bar{\tau}}{4}\left[\bar{\omega}e^{\frac i2
                  (\sqrt{\tau}z + \sqrt{\bar{\tau}}\bar z)} + \omega
                  e^{-\frac i2(\sqrt{\tau}z + \sqrt{\bar{\tau}}\bar z)}
                  - ue^{\frac i2(\sqrt{\tau}z - \sqrt{\bar{\tau}}\bar z)}
                  - ve^{-\frac i2(\sqrt{\tau}z - \sqrt{\bar{\tau}}\bar z)}
                  \right]^{-2}\,,
\end{equation}

where $\omega \in \bC$ and $u,v \in \bR$ denote arbitrary constants
satisfying $\omega\bar{\omega} - uv = |\tau|/4$\footnote{This equality comes
from the requirement that the group element $g \in$ SL$(2,\bC)$
that appears in the WZNW action, and whose Gauss decomposition contains
$\Phi$ \cite{Cacciatori:2001un}, must have unit determinant.}.
If we set $\omega = \rho \sqrt{|\tau|} \exp(i\zeta)$, $\rho \ge 0$,
and translate

\begin{equation}
\sqrt{\tau}z \to \sqrt{\tau}z + \zeta + i\ln\frac{u}{\sqrt{|\tau|}(r -
                 \frac 12)}\,,
\end{equation}

we obtain

\begin{eqnarray}
e^{\gamma\Phi} &=&  \frac{|\tau|}{4}\left[\rho e^{\frac i2
                    (\sqrt{\tau}z + \sqrt{\bar{\tau}}\bar z)} + \rho
                    e^{-\frac i2(\sqrt{\tau}z + \sqrt{\bar{\tau}}\bar z)}
                    \right. \nonumber \\
               & &  \qquad - \left.(\rho - \frac 12) e^{\frac i2(\sqrt{\tau}z -
                    \sqrt{\bar{\tau}}\bar z)}
                    - (\rho + \frac 12) e^{-\frac i2(\sqrt{\tau}z -
                    \sqrt{\bar{\tau}}\bar z)}\right]^{-2}\,.
                    \label{LiouvillesolJ}
\end{eqnarray}

Let us now specify to the case $J=0$ and $\mu \ge 0$.
We choose $\rho = 0$ in (\ref{LiouvillesolJ}). This yields
the $\phi$-independent solution

\begin{equation}
e^{\gamma\Phi} = \mu\left[e^{\frac i2 \sqrt{\mu} (z - \bar z)} -
                 e^{-\frac i2 \sqrt{\mu} (z - \bar z)}\right]^{-2}\,.
\end{equation}

Now remember that the asymptotic boundary of the spacetime (\ref{KdS})
has the topology of a cylinder. One can pass to the plane (with
coordinates $w, \bar w$) by the transformation (\ref{cylplane}),
leading to

\begin{equation}
e^{\gamma\Phi}\,dz d\bar z = \frac{\mu}{(w\bar w)^{1 - \sqrt{\mu}}\left[
                           1 - (w\bar w)^{\sqrt{\mu}}\right]^2}\, dw d\bar w\,,
\label{elliptic}
\end{equation}

which is the standard classical elliptic solution of Liouville
theory \cite{Seiberg:1990eb}.

Semiclassically, the Liouville vertex operators $e^{\alpha\Phi}$ appear
as sources of curvature in the classical equation of motion and lead
to solutions with local elliptic monodromy with $\sqrt{\mu} = 1 - \gamma
\alpha$ \cite{Seiberg:1990eb}. From this we obtain

\begin{equation}
\alpha = \frac{1 - \sqrt{\mu}}{\gamma}\,,
\label{relalphamu}
\end{equation}

i.~e.~, a relation between the mass parameter $\mu$ of the dS$_3$
solution and the parameter $\alpha$ of the vertex operator\footnote{In
the AdS$_3$ case, an interpretation of Liouville non-normalizable
states in terms of particles moving in the bulk was developed
in \cite{Krasnov:2000ia}.}.
In the classical theory, $e^{\alpha\Phi}$ has conformal dimension

\begin{equation}
\Delta_{\mathrm{class}}(e^{\alpha\Phi}) = \frac{\alpha}{\gamma}
= \frac{1 - \sqrt{\mu}}{\gamma^2} = \frac{l}{8G}(1 - \sqrt{\mu})\,.
\label{confdimclass}
\end{equation}

Now the Schwarzschild-de~Sitter solution contains a pair of conical
defects at antipodal points on the spatial two-sphere, so only one of
these is inside the cosmological horizon. From the Gauss-Bonnet
theorem applied to the upper hemisphere containing the point $r=0$, we
obtain

\[
\int R^{(2)} = 4\pi(1-\sqrt{\mu}) + C\,,
\]

where $R^{(2)}$ denotes the spatial curvature scalar and $C$ is a
constant (the integral of $R^{(2)}$ over the space as if there were no
defect). This means that there is a curvature singularity with strength 

\[
R^{(2)}_{\mathrm{sing}}=4\pi (g^{(2)})^{-1/2}(1-\sqrt{\mu})\delta(\vec{r})\,.
\]
 
The Hamiltonian constraint for a static solution is\footnote{The stress
tensor appearing in Eq.~(\ref{Hamconstr}) is not to be confused with
(\ref{stresstensor}).}

\begin{equation}
R^{(2)}-2\Lambda = 16\pi GT^{00}\,, \label{Hamconstr}
\end{equation}

so by comparison we get

\[
\sqrt{g^{(2)}}\,T^{00} = \frac{1}{4G}(1-\sqrt{\mu})\delta(\vec{r})\,.
\]

In general, the stress-energy tensor for pointlike masses located at
isolated points $\vec{r}_i$ is
$\sqrt{g^{(2)}}\,T^{00}=\sum_im_i\delta(\vec{r}-\vec{r}_i)$, so we 
learn that the bare mass of the conical defect is

\begin{equation}
m_{\mathrm{class}} = \frac{1-\sqrt{\mu}}{4G}\,.
\end{equation}

Comparing this with (\ref{confdimclass}), one obtains

\begin{equation}
lm_{\mathrm{class}} = \Delta_{\mathrm{class}} +
                      \bar{\Delta}_{\mathrm{class}}\,,
\end{equation}

so the classical conformal weights reproduce exactly the mass of the
classical point particle in dS$_3$ that causes the conical defect.

The quantum dimension of the operator $e^{\alpha\Phi}$ is given by

\begin{equation}
\Delta(e^{\alpha\Phi}) = \frac{\alpha}{\gamma} - \frac 12 \alpha^2
                         + {\cal O}(1) = \frac{1 - \mu}{2\gamma^2}
                         = \frac{c}{24}(1 - \mu) = \frac{l}{16G}(1 - \mu)\,.
\end{equation}

Using $\bar{\Delta} = \Delta$ (the vertex operators are scalars),
and $\tilde c = c$, we obtain

\begin{equation}
\Delta + \bar{\Delta} = lM + \frac{c + \tilde c}{24}\,,
\end{equation}

so that $(\Delta + \bar{\Delta})/l$ coincides (modulo a constant shift
of $(c+\tilde c)/24l$ coming from the transformation (\ref{cylplane}) from
the cylinder to the plane) with the mass $M$ that we computed in
(\ref{MJ}). 

This mass includes the contribution of the gravitational field, and
should not be confused with the bare mass of
the particles that cause the conical defects and which act as static
external conditions. In particular, since $\mu$ must be strictly
positive to have a static solution, the bare masses should be bounded
from above, a fact that will be confirmed below from the Seiberg
bound. If we like, we can think of the mass that it is a kind of
gravitational dressing effect. 

We have thus established a one to one correspondence between Liouville
vertex operators in the boundary CFT and de~Sitter solutions in the bulk.
The quantum conformal weights of the operators are identical to
the masses of the bulk solutions. Also, as already noticed above,
the spectrum of conformal dimensions in Liouville theory, unlike the one
of a generic CFT, is not bounded from below. Instead one has
$\Delta \le Q^2/8$ \cite{Seiberg:1990eb}, where $Q$ denotes the
background charge\footnote{Strictly speaking, in the
quantum theory one has $Q = 2/\gamma + \gamma$. As $\gamma^2 = 8l_P/l$
($l_P = G$ denotes the Planck length in three dimensions), and
we only consider the regime $l \gg l_P$ where one can trust classical gravity,
we have $\gamma \ll 1$, so that the
corrections to the classical background charge $2/\gamma$ can be neglected.},
$Q = 2/\gamma$. This leads to the inequality
$\mu \ge 0$, but otherwise $\mu$ is unrestricted. In particular,
values $\mu > 1$ corresponding to excess angles, yield $\Delta < 0$, and
thus seem to be entirely consistent from the Liouville point of view.
Furthermore, we note
that the solution with $\mu = 0$, which is essentially de~Sitter space
in inflationary coordinates, has $\alpha = 1/\gamma$, i.~e.~, it
corresponds to the puncture operator.

A further point to check is the equality of the Liouville stress tensor
and the Brown-York energy-momentum tensor (\ref{stresstensw}) of
three-dimensional de~Sitter gravity. For $J = 0$, the latter is given by

\begin{equation}
T(w) = \frac{1 - \mu}{4\pi\gamma^2 w^2}\,, \qquad
\overline{T}(\bar w) = \frac{1 - \mu}{4\pi\gamma^2 \bar w^2}\,,
\end{equation}

which indeed coincides with the Liouville stress tensor for elliptic
solutions \cite{Seiberg:1990eb}. Note also that $T(w)$ can be written as

\begin{equation}
T(w) = \frac{1}{2\pi\gamma^2}S[f(w); w]\,,
\end{equation}

where

\begin{equation}
S[f(w); w] = \frac{f'''}{f'} - \frac 32 \left(\frac{f''}{f'}\right)^2
\end{equation}

denotes the Schwartzian derivative, and the "uniformizing map" $f(w)$
is given by $f(w) = w^{\sqrt{\mu}}$.

Let us now consider the gravity solutions with $\mu < 0$.
We will see that they also have a nice interpretation in
Liouville theory.

Defining $E$ by $\sqrt{\mu} = iE\gamma$,
one gets for the classical Liouville solution corresponding to (\ref{KdS})
with $J=0$ and $\mu < 0$\footnote{It is interesting to note that
the solutions with temperature ($\mu > 0$) have elliptic monodromy,
whereas those without temperature ($\mu < 0$) have hyperbolic
monodromy. In AdS$_3$ it happens the other way round:
The BTZ black hole has hyperbolic monodromy, whereas a particle in
AdS$_3$ has elliptic monodromy \cite{Martinec:1998wm}.}

\begin{equation}
e^{\gamma\Phi} dzd\bar z = \frac{E^2\gamma^2}{4w\bar w\sin^2\left(
                           \frac{E\gamma}{2}\ln w\bar w\right)}dwd\bar w\,,
\label{hyperbolic}
\end{equation}

which can also be obtained from (\ref{elliptic}) by analytical continuation.
As is well-known (cf.~e.~g.~\cite{Ginsparg:is} for a review), the hyperbolic
solution (\ref{hyperbolic}) corresponds in the semiclassical limit
$\gamma \to 0$ to the normalizable quantum states $\psi_E$ with momentum $E$,
i.~e.~, to the so-called macroscopic states. Formally, these states
can be associated to vertex operators $e^{\alpha\Phi}$ with

\begin{equation}
\alpha = \frac Q2 + iE\,. \label{alphaQE}
\end{equation}

If we use $\sqrt{\mu} = iE\gamma$ in our relation (\ref{relalphamu})
that connects vertex operators and bulk solutions, we get exactly
(\ref{alphaQE}). Furthermore, since the quantum state $\psi_E$ has
energy $E^2/2 + Q^2/8$ \cite{Seiberg:1990eb}, the sum of the energies
of $\psi_E$ (right-moving) and $\psi_{-E}$ (left-moving) is

\begin{equation}
E^2 + \frac{Q^2}{4} = \frac{1 - \mu}{\gamma^2} = lM +
\frac{c + \tilde c}{24}\,,
\end{equation}

which gives again the mass (\ref{MJ}).

Summarizing, one has thus the following picture for $J=0$: Gravity solutions
that have a temperature (i.~e.~, with $\mu \ge 0$) correspond to
vertex operators with $\alpha = (1 - \sqrt{\mu})/\gamma$, i.~e.~,
to non-normalizable or microscopic states.
Solutions with $\mu < 0$ correspond to normalizable or macroscopic
states with real momentum $E$, where $E$ is given by $\sqrt{\mu} = iE\gamma$.
Also the Seiberg bound \cite{Seiberg:1990eb}, which states that operators
with $\alpha > Q/2$ do not exist, has an interpretation in de~Sitter
gravity. Using (\ref{confdimclass}) and $lm_{\mathrm{class}} =
\Delta_{\mathrm{class}} + \bar\Delta_{\mathrm{class}}$, the Seiberg bound
implies an upper mass bound $m_{\mathrm{class}} \le 1/4G$ for classical
point particles in de~Sitter space.

Up to now we only considered the case of vanishing angular
momentum. It is of course natural to ask how to interpret the rotating
Kerr-dS$_3$ solution on the CFT side. The first point we note is
that one can formally associate the solution to CFT operators
with complex conformal weights

\begin{equation}
\Delta = \frac{1 - \tau}{2\gamma^2}\,, \qquad
\bar{\Delta} = \frac{1 - \bar{\tau}}{2\gamma^2}\,, \label{confdimJ}
\end{equation}

in terms of which

\begin{equation}
lM = \Delta + \bar{\Delta} - \frac{c + \tilde c}{24}\,, \qquad
iJ = \Delta - \bar{\Delta}\,.
\end{equation}

In Liouville theory, one may consider more general vertex operators
$e^{\alpha\Phi}$ with $\alpha \in \bC$ \cite{Teschner:2001rv}. These are
in one-to-one correspondence with microscopic states if and only if
$\Re(\alpha) < Q/2$ \cite{Teschner:2001rv}, which is the generalized Seiberg
bound. If we replace $\mu$ by $\tau$ in (\ref{relalphamu}) for $J \neq 0$,
we get for the case of nonvanishing angular momentum

\begin{equation}
\alpha = \frac{1 - \sqrt{\tau}}{\gamma}\,, \label{relalphatau}
\end{equation}

which is complex and has $\Re(\alpha) \neq Q/2$, i.~e.~, does not
correspond to normalizable states. Using (\ref{tau}), one obtains from
(\ref{relalphatau})

\begin{equation}
\alpha = \frac{1}{\gamma} \mp \sqrt{\frac{\mu + |\tau|}{2\gamma^2}} \pm
         \frac{i\gamma J}{\sqrt{2(\mu + |\tau|)}}\,.
         \label{alphawithJ}
\end{equation}

If we choose the upper sign in (\ref{alphawithJ}), the Seiberg bound
$\Re(\alpha) < 1/\gamma$ is satisfied. One can thus make sense of
vertex operators $e^{\alpha\Phi}$, with $\alpha$ given by the upper sign
expression in (\ref{alphawithJ}). In the quantum theory, these have
conformal weights \cite{Teschner:2001rv}

\begin{equation}
\Delta = \frac{\alpha}{\gamma} - \frac 12 \alpha^2 = \frac{1 - \mu}
         {2\gamma^2} + \frac{iJ}{2}\,,
\end{equation}

which is exactly (\ref{confdimJ}). In conclusion, we showed that the
Kerr-de~Sitter solution with nonvanishing angular momentum can be associated
to vertex operators $e^{\alpha\Phi}$ with $\alpha$ as in (\ref{alphawithJ})
with the upper signs.

\section{Is the dual CFT thermal?}

\label{thermal}

Recently, some criticisms concerning the existence of a dS/CFT
correspondence appeared in the literature \cite{Dyson:2002nt}.
In particular, the authors of \cite{Dyson:2002nt} considered
a general finite closed system described by a thermal density
matrix, and a thermal correlator

\begin{equation}
F(t) = <{\cal O}(0) {\cal O}(t)>\,.
\end{equation}

It was then shown that the long time average of $F(t)F^{\ast}(t)$
is non-zero and positive, which leads to a contradiction with the dS/CFT
result in static coordinates \cite{Klemm:2001ea}\footnote{We only consider
the simple case of operators ${\cal O}$ that couple to bulk scalars
of mass $m$.},

\begin{equation}
<{\cal O}(0,0) {\cal O}(t,\phi)> \sim \left[\cosh\frac tl - \cos\phi
\right]^{-h}\,, \label{cylcorr}
\end{equation}

where $h = 1 + \sqrt{1 - m^2l^2}$. (\ref{cylcorr}) is not the
standard thermal correlator, rather it is the two-point function
for dimension $(h,h)$ operators on a cylinder, whose length
(not circumference) is parametrized by the Euclidean time coordinate
$t$. (\ref{cylcorr}) behaves like

\begin{equation}
<{\cal O}(0,0) {\cal O}(t,\phi)> \sim e^{-ht/l} \label{asbehav}
\end{equation}

for large $t$.
Clearly the behaviour (\ref{asbehav}) would imply a zero
long-time average. Obviously the apparent contradiction
found in \cite{Dyson:2002nt}, which is based on the assumption that the
dual CFT is described by a thermal density matrix,
is resolved if the conformal field theory does
not encode the thermal nature of de~Sitter space. We would like to
point out here some arguments in favour of this.

First of all, the concept of assigning a temperature to de~Sitter
space is well-defined only in the static patches. However, the
past (and future-) boundary, where the CFT resides, lies outside the
static region.
In particular, the local Tolman temperature of de~Sitter space,

\begin{equation}
T(r) = \frac{1}{2\pi l\sqrt{1 - \frac{r^2}{l^2}}}
\end{equation}

formally becomes imaginary for $r>l$. It might thus be that
the conformal field theory on ${\cal I}^-$ does not capture
the thermal nature of de~Sitter space.
If this is true, and if the dual CFT nevertheless
accounts somehow for de~Sitter entropy, we do not expect this
entropy to be thermal\footnote{A non-thermal interpretation of de~Sitter
entropy in terms of a sort of Euclidean entanglement entropy was recently
proposed in \cite{Kabat:2002hj}.}.

In section \ref{KdSsol} we argued that the partition
function of dS gravity might have a non-thermal interpretation. We
will now confirm that in our Liouville approach this is indeed the
case. Let us start with the KPZ equation for gravitational dressing
of CFT operators with bare conformal weight $\Delta_0$ by vertex operators
$e^{\alpha\Phi}$,

\begin{equation}
\alpha - \frac Q2 = -\sqrt{\frac 14 Q^2 - 2 + 2\Delta_0}\,.
\end{equation}

Setting $\Delta = 1 - \Delta_0$ yields

\begin{equation}
\alpha - \frac Q2 = -\sqrt{\frac 14 Q^2 - 2\Delta}\,,
\label{KPZDelta}
\end{equation}

which is of course the formula for the quantum conformal weights
$\Delta$ of vertex operators $e^{\alpha\Phi}$. Now observe that the
entropy of the Schwarz\-schild-dS$_3$ solution ((\ref{KdS}) with $J=0$)
is given by

\begin{equation}
S = \frac{\pi l\sqrt{\mu}}{2G} = \frac{4\pi \sqrt{\mu}}{\gamma^2}\,,
\end{equation}

and, using (\ref{relalphamu}), that

\begin{equation}
\alpha - \frac Q2 = -\frac{\sqrt{\mu}}{\gamma} = -\frac{S\gamma}{4\pi}\,.
\label{alphaQS}
\end{equation}

Inserting (\ref{alphaQS}) into (\ref{KPZDelta}), one obtains

\begin{equation}
S = 2\pi\sqrt{2Q^2\left(\frac{Q^2}{8} - \Delta\right)}\,.
\end{equation}

If we finally use the central charge $c = 3Q^2$ and $\bar{\Delta} = \Delta$,
we get

\begin{equation}
S = 2\pi\sqrt{\frac c6\left(\frac{c}{24} - \Delta\right)} +
    2\pi\sqrt{\frac c6\left(\frac{c}{24} - \bar\Delta\right)}\,.
\label{Cardy}
\end{equation}

(\ref{Cardy}) looks like the Cardy formula for the asymptotic level
density of conformal field theories, but actually it is not, because
the signs of the terms $\Delta$ and $c/24$ are interchanged. Rather,
we saw that (\ref{Cardy}) is the KPZ equation. Looking at (\ref{alphaQS}),
one also sees what de~Sitter entropy corresponds to in Liouville
theory. Using

\begin{equation}
iE = \alpha - \frac Q2
\end{equation}

for the Liouville momentum $E$, one finally obtains

\begin{equation}
S = -\frac{4\pi i}{\gamma}E\,,
\label{SE}
\end{equation}

so that de~Sitter entropy is essentially Liouville momentum.
In particular, it has no statistical meaning in this approach.
Note that the Schwarz\-schild-dS$_3$ solution with cosmological
horizon corresponds to imaginary momentum $E$, so that $S$ is real,
as it should be.

\section{Final remarks}

In summary, we established a detailed and quantitative correspondence
between three-dimensional gravity with positive cosmological constant
and Euclidean Liouville theory.
Of course the relationship between gravity in three dimensions
and Liouville theory at the classical level is rather well-known;
in fact the appearance of the Liouville equation in the context of
three-dimensional de~Sitter gravity was already pointed out
in \cite{Deser:dr}. What is less obvious, and comes as a surprise,
is that classical gravity with positive cosmological constant encodes
the quantum properties of Liouville theory. Indeed we found that the
quantum dimensions of Liouville vertex operators coincide exactly
with the masses of the gravity solutions, computed with the Brown-York
stress tensor, which is an entirely classical concept. This fits
nicely into the usual notion of duality: On one side, we are at
a classical level, whereas in the dual theory we are in the quantum
regime.

A further point that seems worth mentioning is that all the
bulk solutions with cosmological event horizons (i.~e.~, the ones
with $J \neq 0$ or with $J = 0$, $\mu > 0$ in (\ref{KdS})) correspond
to microscopic states in Liouville theory, whereas the solutions without
horizons ($J = 0$, $\mu < 0$) correspond to macroscopic (normalizable)
states. This may be a pure coincidence, but it would certainly be
interesting to understand this better. We note in this context that
solutions with event horizon have a (real) entropy $S$, and thus,
since $S$ is related to the Liouville momentum $E$ by
(\ref{SE})\footnote{(\ref{SE}) is valid for $J=0$, but probably there
exists an appropriate generalization to the spinning case.},
must have imaginary $E$, i.~e.~, must correspond to non-normalizable
(microscopic) states.

We also note that turning on angular momentum on the bulk side
leads to complex conformal weights, and we saw that such conformal
weights indeed naturally appear in Liouville theory. This is what
one expects from a CFT that is dual to dS gravity.

It would be interesting to check if the Liouville partition function
(probably with vertex operator insertions) reproduces the
gravitational (bulk) partition function. This is currently under
investigation.

\section*{Acknowledgements}
\small

This work was partially supported by INFN, MURST and
by the European Commission RTN program
HPRN-CT-2000-00131, in which D.~K.~is
associated to the University of Torino.
The authors would like to thank Sergio Cacciatori for helpful discussions,
and Benedicte Ponsot for clarifying correspondence.


\normalsize



\newpage

\begin{thebibliography}{99}

\bibitem{Balasubramanian:2001rb}
V.~Balasubramanian, P.~Horava and D.~Minic,
JHEP {\bf 0105} (2001) 043
[arXiv:hep-th/0103171].

\bibitem{Witten:2001kn}
E.~Witten,
arXiv:hep-th/0106109.

\bibitem{Gibbons:mu}
G.~W.~Gibbons and S.~W.~Hawking,
Phys.\ Rev.\ D {\bf 15} (1977) 2738.

\bibitem{Maldacena:1998ih}
J.~Maldacena and A.~Strominger,
JHEP {\bf 9802} (1998) 014
[arXiv:gr-qc/9801096].

\bibitem{Park:1998qk}
M.~I.~Park,
Phys.\ Lett.\ B {\bf 440} (1998) 275
[arXiv:hep-th/9806119].

\bibitem{Banados:1998tb}
M.~Ba\~{n}ados, T.~Brotz and M.~E.~Ortiz,
Phys.\ Rev.\ D {\bf 59} (1999) 046002
[arXiv:hep-th/9807216].

\bibitem{Kim:1998zs}
W.~T.~Kim,
Phys.\ Rev.\ D {\bf 59} (1999) 047503
[arXiv:hep-th/9810169].

\bibitem{Lin:1999gf}
F.~L.~Lin and Y.~S.~Wu,
Phys.\ Lett.\ B {\bf 453} (1999) 222
[arXiv:hep-th/9901147].

\bibitem{Hawking:2000da}
S.~Hawking, J.~Maldacena and A.~Strominger,
JHEP {\bf 0105} (2001) 001
[arXiv:hep-th/0002145].

\bibitem{Govindarajan:2002ry}
T.~R.~Govindarajan, R.~K.~Kaul and V.~Suneeta,
arXiv:hep-th/0203219.

\bibitem{'tHooft:gx}
G.~'t Hooft,
arXiv:gr-qc/9310026;\\
L.~Susskind,
J.\ Math.\ Phys.\  {\bf 36} (1995) 6377
[arXiv:hep-th/9409089].

\bibitem{Aharony:1999ti}
O.~Aharony, S.~S.~Gubser, J.~Maldacena, H.~Ooguri and Y.~Oz,
Phys.\ Rept.\  {\bf 323} (2000) 183
[arXiv:hep-th/9905111].

\bibitem{Hull:1998vg}
C.~M.~Hull,
JHEP {\bf 9807} (1998) 021
[arXiv:hep-th/9806146].

\bibitem{Bousso:1999cb}
R.~Bousso,
JHEP {\bf 9906} (1999) 028
[arXiv:hep-th/9906022].

\bibitem{Strominger:2001pn}
A.~Strominger,
JHEP {\bf 0110} (2001) 034
[arXiv:hep-th/0106113].

\bibitem{Klemm:2001ea}
D.~Klemm,
Nucl.\ Phys.\ B {\bf 625} (2002) 295
[arXiv:hep-th/0106247].

\bibitem{Cacciatori:2001un}
S.~Cacciatori and D.~Klemm,
Class.\ Quant.\ Grav.\  {\bf 19} (2002) 579
[arXiv:hep-th/0110031].

\bibitem{Balasubramanian:2001nb}
V.~Balasubramanian, J.~de Boer and D.~Minic,
arXiv:hep-th/0110108.

\bibitem{Bousso:2001mw}
R.~Bousso, A.~Maloney and A.~Strominger,
arXiv:hep-th/0112218.

\bibitem{Spradlin:2001nb}
M.~Spradlin and A.~Volovich,
arXiv:hep-th/0112223.

\bibitem{Leblond:2002ns}
F.~Leblond, D.~Marolf and R.~C.~Myers,
arXiv:hep-th/0202094.

\bibitem{Kabat:2002hj}
D.~Kabat and G.~Lifschytz,
arXiv:hep-th/0203083.

\bibitem{Mazur:2001aa}
P.~O.~Mazur and E.~Mottola,
Phys.\ Rev.\ D {\bf 64} (2001) 104022
[arXiv:hep-th/0106151];
M.~Li,
arXiv:hep-th/0106184;
S.~Nojiri and S.~D.~Odintsov,
Phys.\ Lett.\ B {\bf 519} (2001) 145
[arXiv:hep-th/0106191];
A.~Chamblin and N.~D.~Lambert,
Phys.\ Rev.\ D {\bf 65} (2002) 066002
[arXiv:hep-th/0107031];
Y.~H.~Gao,
arXiv:hep-th/0107067;
J.~Bros, H.~Epstein and U.~Moschella,
arXiv:hep-th/0107091;
E.~Halyo,
arXiv:hep-th/0107169;
R.~Kallosh,
arXiv:hep-th/0109168;
Y.~S.~Myung,
Mod.\ Phys.\ Lett.\ A {\bf 16} (2001) 2353
[arXiv:hep-th/0110123];
B.~G.~Carneiro da Cunha,
arXiv:hep-th/0110169;
R.~G.~Cai, Y.~S.~Myung and Y.~Z.~Zhang,
arXiv:hep-th/0110234;
U.~H.~Danielsson,
arXiv:hep-th/0110265;
A.~C.~Petkou and G.~Siopsis,
JHEP {\bf 0202} (2002) 045
[arXiv:hep-th/0111085];
R.~G.~Cai,
Phys.\ Lett.\ B {\bf 525} (2002) 331
[arXiv:hep-th/0111093];
arXiv:hep-th/0112253;
M.~Li and F.~L.~Lin,
arXiv:hep-th/0111201;
A.~M.~Ghezelbash and R.~B.~Mann,
JHEP {\bf 0201} (2002) 005
[arXiv:hep-th/0111217];
M.~H.~Deh\-ghani,
arXiv:hep-th/0112002;
M.~Cveti\v{c}, S.~Nojiri and S.~D.~Odintsov,
arXiv:hep-th/0112045;
A.~J.~Medved,
arXiv:hep-th/0112226;
S.~R.~Das,
arXiv:hep-th/0202008;
F.~Larsen, J.~P.~van der Schaar and R.~G.~Leigh,
arXiv:hep-th/0202127;
G.~Siopsis,
arXiv:hep-th/0203208.

\bibitem{Tagirov:1972vv}
E.~A.~Tagirov,
Annals Phys.\  {\bf 76} (1973) 561.

\bibitem{Hull:2001ii}
C.~M.~Hull,
JHEP {\bf 0111} (2001) 012
[arXiv:hep-th/0109213].

\bibitem{Seiberg:1990eb}
N.~Seiberg,
Prog.\ Theor.\ Phys.\ Suppl.\  {\bf 102} (1990) 319.

\bibitem{Dyson:2002nt}
L.~Dyson, J.~Lindesay and L.~Susskind,
arXiv:hep-th/0202163.

\bibitem{Brown:1992br}
J.~D.~Brown and J.~W.~York,
Phys.\ Rev.\ D {\bf 47} (1993) 1407.

\bibitem{Teitelboim:2002cv}
C.~Teitelboim,
arXiv:hep-th/0203258.

\bibitem{Weinberg:1988cp}
S.~Weinberg,
Rev.\ Mod.\ Phys.\  {\bf 61} (1989) 1.

\bibitem{Martinec:1984fs}
E.~J.~Martinec,
Phys.\ Rev.\ D {\bf 30} (1984) 1198.

\bibitem{Witten:1988hc}
E.~Witten,
Nucl.\ Phys.\ B {\bf 311} (1988) 46.

\bibitem{Moncrief:1989dx}
V.~Moncrief,
J.\ Math.\ Phys.\  {\bf 30} (1989) 2907.

\bibitem{Ezawa:1993ti}
K.~Ezawa,
Phys.\ Rev.\ D {\bf 49} (1994) 5211
[Addendum-ibid.\ D {\bf 50} (1994) 2935]
[arXiv:hep-th/9311103].

\bibitem{Elitzur:1989nr}
S.~Elitzur, G.~W.~Moore, A.~Schwimmer and N.~Seiberg,
Nucl.\ Phys.\ B {\bf 326} (1989) 108.

\bibitem{Krasnov:2000ia}
K.~Krasnov,
Class.\ Quant.\ Grav.\  {\bf 18} (2001) 1291
[arXiv:hep-th/0008253].

\bibitem{Ginsparg:is}
P.~Ginsparg and G.~W.~Moore,
arXiv:hep-th/9304011.

\bibitem{Martinec:1998wm}
E.~J.~Martinec,
arXiv:hep-th/9809021.

\bibitem{Teschner:2001rv}
J.~Teschner,
Class.\ Quant.\ Grav.\  {\bf 18} (2001) R153
[arXiv:hep-th/0104158].

\bibitem{Deser:dr}
S.~Deser and R.~Jackiw,
Annals Phys.\  {\bf 153} (1984) 405.

\end{thebibliography}
\end{document}